# Anomalous Spin-Orbit Torques in Magnetic Single-Layer Films

Wenrui Wang[1*], Tao Wang[2*], Vivek P. Amin[3,4], Yang Wang[2], Anil Radhakrishnan[1], Angie Davidson[5], Shane R. Allen[5], T. J. Silva[6], Hendrik Ohldag[7], Davor Balzar[5], Barry L. Zink[5], Paul M. Haney[4], John Q. Xiao[2], David G. Cahill[8], Virginia O. Lorenz[1], Xin Fan[5]

1. Department of Physics, University of Illinois at Urbana-Champaign, Urbana, Illinois 61801, USA
2. Department of Physics and Astronomy, University of Delaware, Newark, DE 19716, USA
3. Maryland Nanocenter, University of Maryland, College Park, MD 20742, USA
4. Center for Nanoscale Science and Technology, National Institute of Standards and Technology, Gaithersburg, Maryland 20899, USA
5. Department of Physics and Astronomy, University of Denver, Denver, CO 80210, USA
6. Quantum Electromagnetics Division, National Institute of Standards and Technology, Boulder, CO, 80305, USA
7. Stanford Synchrotron Radiation Lightsource, SLAC National Accelerator Laboratory, Menlo Park, California 94025, USA
8. Department of Materials Science and Engineering, University of Illinois at Urbana-Champaign, Urbana, Illinois 61801, USA

vlorenz@illinois.edu

xin.fan@du.edu

* These two authors contributed equally to this paper

**Abstract:**

Spin-orbit interaction (SOI) couples charge and spin transport[1-3], enabling electrical control of magnetization[4,5]. A quintessential example of SOI-induced transport is the anomalous Hall effect (AHE)[6], first observed in 1880, in which an electric current perpendicular to the magnetization in a magnetic film generates charge accumulation on the surfaces. Here we report the observation of a counterpart of the AHE that we term the anomalous spin-orbit torque (ASOT), wherein an electric current parallel to the magnetization generates opposite spin-orbit torques on the surfaces of the magnetic film. We interpret the ASOT as due to a spin-Hall-like current generated with an efficiency of $0.053 \pm 0.003$ in $Ni_{80}Fe_{20}$, comparable to the spin Hall angle of Pt[7]. Similar effects are also observed in other common ferromagnetic metals, including Co, Ni, and Fe. First principles calculations corroborate the order of magnitude of the measured values. This work suggests that a strong spin current with spin polarization transverse to magnetization can exist in a ferromagnet, despite spin dephasing[8]. It challenges the current understanding of spin-orbit torque in magnetic/nonmagnetic bilayers, in which the charge-spin conversion in the magnetic layer has been largely neglected.

Spin-orbit interaction can convert a charge current into a flow of spin angular momentum (spin current) with spin polarization orthogonal to both the charge and spin currents[9]. One of its manifestations in a magnetic conductor is the AHE[10], illustrated in Fig. 1a. Due to the imbalance of electrons with spins parallel and antiparallel to the magnetization, the flow of spin current results in charge accumulation on the top and bottom surfaces. The spin current in this configuration is polarized parallel with the magnetization[11-13]. Applying similar considerations to the ASOT configuration, illustrated in Fig. 1b, in which the electric current is parallel to the magnetization, SOI should also give rise to a spin current flowing between the top and bottom surfaces of the magnetic conductor, except with electron spins transverse to the magnetization. The existence of transversely polarized spin current in a ferromagnet may sound counterintuitive, because transverse spins precess rapidly about the magnetization and are subject to strong dephasing[8,14]. However, recently it has been theoretically predicted that transversely polarized spin current is allowed in diffusive ferromagnets[15] because the spin-orbit interaction, which generates spin current, competes with spin dephasing. In this paper, we also show that transversely polarized spin current can exist in ferromagnets in the clean limit, using first-principles calculations. We refer to the mechanism of the current-induced transversely polarized spin current as the transverse spin Hall effect (TSHE). We emphasize that the TSHE is fundamentally different from the previously studied spin current generation in the AHE configuration, where the spin polarization is necessarily parallel with the magnetization.

The transversely polarized spin current does not give rise to a bulk spin torque, due to symmetry, in ferromagnets with bulk inversion symmetry. Instead, we predict that it will result in net ASOT in the $y$-direction on the top and bottom surfaces, where inversion symmetry is broken. (see Supplementary Information section S1) Under the assumption that ASOT results in a small perturbation to the magnetization, the ASOTs are equivalent to effective magnetic fields in the $z$-direction[16] that tilt the magnetization out of plane, as illustrated in Fig. 1b. Like the AHE, ASOT is a fundamental property of all ferromagnetic and ferrimagnetic conductors (those with broken bulk inversion symmetry have been shown to exhibit a non-zero bulk spin-orbit torque[17,18]).

The out-of-plane magnetization tilting, $m_z^{ASOT}$, due to the ASOT at the top ($\tau_T^{ASOT}$) and bottom ($\tau_B^{ASOT}$) surfaces can be derived as

$$m_z^{ASOT}(z) = \frac{\tau_T^{ASOT}\cosh\frac{d-z}{\lambda} + \tau_B^{ASOT}\cosh\frac{z}{\lambda}}{\lambda\sinh\frac{d}{\lambda}(|H_{ext}|+M_{eff})\mu_0 M_S} m_x \quad (1)$$

where $d$ is the total thickness of the film, $\lambda$ is the exchange length, $H_{ext}$ is an applied external magnetic field in the $x$-direction, $M_{eff}$ is the effective demagnetizing field, $M_s$ is the saturation magnetization, and $m_x$ is the projection of the unit magnetization along the $x$-direction. Here, the ASOT is assumed to be located only at the surfaces and the surface anisotropy is neglected. See Supplementary Information section S4 for the derivation of Eq. (1), a discussion of why ASOT can be treated as a pure surface effect, and a numerical analysis that takes into account of the surface anisotropy.

Because the exchange coupling in the magnetic material aligns the magnetization, the spatially-antisymmetric magnetization tilting is expected to be measurable only when the magnetic material is thicker than the exchange length (e.g. 5.1 nm for $Ni_{80}Fe_{20}$). A simulation of the out-of-plane magnetization distribution due to ASOT in a 32 nm $Ni_{80}Fe_{20}$ (Py) film is shown in Fig. 1c.

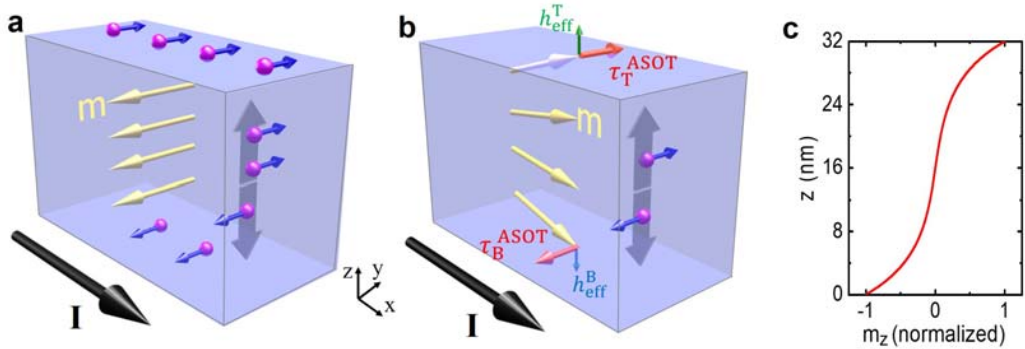

Figure 1 **Illustrations of the anomalous Hall effect and anomalous spin-orbit torque. a**, In the anomalous Hall effect (AHE) a charge current **I** (black arrow) perpendicular to the magnetization **m** (yellow arrows) generates a flow of spin current (grey arrows) in the $z$-direction. Here blue arrows on purple spheres represent spin directions of electrons. Due to the imbalance of majority and minority electrons, the flow of spin current results in spin and charge accumulation on the top and bottom surfaces. **b**, When a charge current is applied parallel with the magnetization, the

AHE vanishes, but spin-orbit interaction generates a flow of transversely polarized spin current that gives rise to anomalous spin-orbit torque (ASOT). The ASOTs (red arrows) are equivalent to out-of-plane fields (green arrows) that tilt the magnetization out of plane. $\tau_T^{ASOT}$ ($\tau_B^{ASOT}$) and $h_{eff}^T$ ($h_{eff}^B$) are the ASOTs and equivalent fields at the top (bottom) surfaces, respectively. **c**, Simulated distribution of the out-of-plane magnetization $m_z$ in a 32 nm Py film driven by equal and opposite ASOTs on the surfaces, scaled by the maximum value.

To observe ASOT, we fabricate a sample with structure substrate/AlO$_x$(2)/Py(32)/AlO$_x$(2)/SiO$_2$(3), where the numbers in parentheses are thicknesses in nanometers; the substrate is fused silica, which allows optical access to the bottom of the sample. Py is chosen because it is magnetically soft and widely used for the study of spin-orbit torques. The film is lithographically patterned into a 50 $\mu$m × 50 $\mu$m square and connected by gold contact pads, as shown in Fig. 2(a). When an electric current $I$ of 40 mA is applied directly through the sample, ASOTs at the top ($\tau_T^{ASOT}$) and bottom ($\tau_B^{ASOT}$) surfaces lead to non-uniform magnetization tilting, as described by Eq. (1). When a calibration current $I_{Cal}$ of 400 mA is passed around the sample, an out-of-plane Oersted field $\mu_0 h_{Cal} \approx 0.85$ mT is generated that uniformly tilts the magnetization out of plane, which is used for calibrating the magnitude of the ASOTs:

$$m_z^{Cal}(z) = \frac{h_{Cal}}{|H_{ext}| + M_{eff}} \tag{2}$$

We detect the magnetization changes using the polar magneto-optic Kerr effect (MOKE) by measuring the Kerr rotation $\theta_k$ and ellipticity change $\varepsilon_k$ of the polarization of a linearly polarized laser reflected from the sample[19,20]. The penetration depth of the laser in Py is approximately 14 nm, which is less than half the thickness of the 32 nm Py. Therefore, the MOKE response is more sensitive to the ASOT-induced out-of-plane magnetization $m_z^{ASOT}(z)$ on the surface on which the laser is directly incident.

The Kerr rotation due to ASOT as a function of the external field (shown in Figs. 2c and d) resembles a magnetization hysteresis, as can be understood from Eq. (1). The overall offsets of the Kerr rotation signals are due to a residual, current-induced out-of-plane Oersted field due to imprecision in locating the MOKE probe spot exactly in the center of the 50 × 50μm² sample, (see Supplementary Information Fig. S4(b) for MOKE signal dependence on the laser spot position), which does not depend on the in-plane magnetization orientation[16]. In contrast, when a uniform calibration field $h_{Cal}$ is applied, the Kerr rotation is symmetric as a function of external field $H_{ext}$ (see Fig. 2e and f), consistent with Eq. (2). The Kerr rotation due to ASOT on the top (Fig. 2c) and bottom (Fig. 2d) surfaces are the same sign, in agreement with our phenomenological model (Fig. 1c), which predicts the bottom ASOT has similar magnitude but opposite sign as the top ASOT. In contrast, the Kerr rotation due to the calibration field (Fig. 2e and f) changes sign because $h_{Cal}$ is reversed upon flipping the sample.

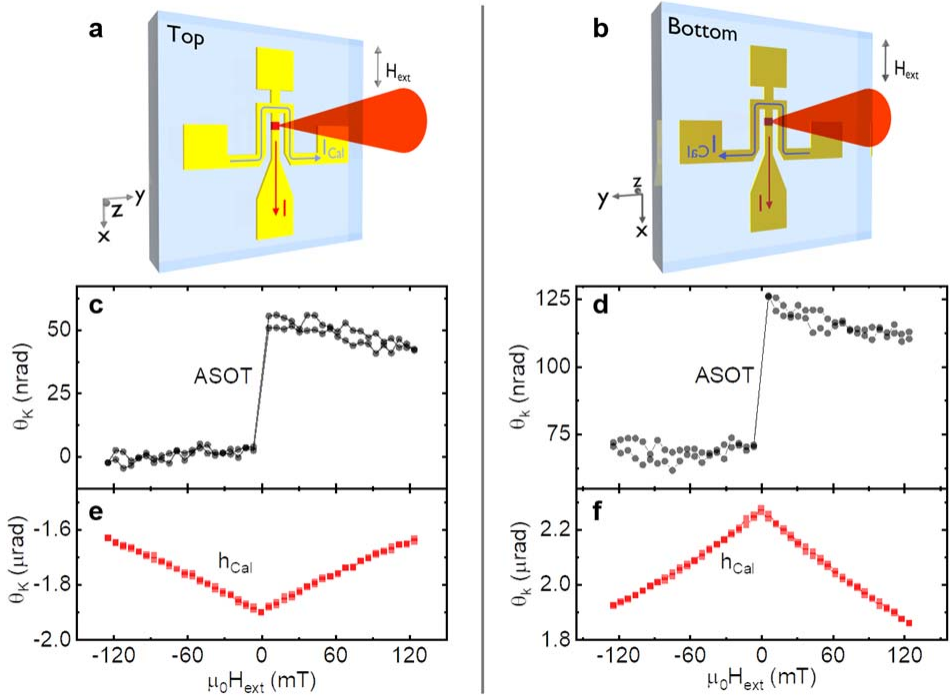

Figure 2 **Symmetry of the anomalous spin-orbit torque.** Diagrams of the measurement configurations with the laser incident on **a**, the top and **b**, the bottom of the sample. The plots below each diagram correspond to signals measured in that diagram's configuration. **c-d**, The measured Kerr rotation signals for when current is applied through the sample, which arise from ASOTs. **e-f**, The measured Kerr rotation signals for when the calibration field $h_{Cal}$ is applied.

As shown in Fig. 3a, the polar MOKE response due to ASOT is linear with applied electric current, indicating no significant heating-related effects up to $5 \times 10^{10}$ A/m$^2$ current density. As shown in Fig. 3b, the polar MOKE response exhibits a cosine dependence on the relative angle between the electric current and the magnetization, consistent with Eq. (1).

Unlike the Oersted field, which depends on the total current, ASOT should depend on the current density. To confirm this, we grow a series of AlO$_x$(2)/Py($t$)/AlO$_x$(2)/SiO$_2$(3) films on silicon substrates with 1 $\mu$m-thick thermal oxide, where $t$ varies from 4 nm to 48 nm. For all samples, we apply the same current density of $5 \times 10^{10}$ A/m$^2$, and use MOKE to quantify the ASOT. To fit the measured MOKE results, we use a propagation matrix method[19] (see method section and Supplementary Information S5) to numerically simulate the MOKE signal as a function of the Py thickness. As presented in Fig. 3c, the validity of the method is first verified by a thickness-dependent calibration measurement, where a uniform 0.85 mT out-of-plane calibration field is applied to all samples. To extract the ASOT amplitude, the top-surface Kerr rotation and the ellipticity change due to the ASOT is fitted in Fig. 3d. The only free fitting parameter is the ASOT on the top surface, $\tau_T^{ASOT}$, which is assumed to be the same for all Py thicknesses under the same current density and to have equal magnitude and opposite sign as the ASOT on the bottom surface $\tau_B^{ASOT}$. The good agreement between experiment and simulation supports the assumption that ASOT depends on current density. The ASOTs are extrapolated to be $\tau_T^{ASOT} = -\tau_B^{ASOT} = (-0.86 \pm 0.04) \times 10^{-6}$ J/m² from the fitting*. Relating this torque to a spin current allows us to find the Spin-Hall-angle-like efficiency of the ASOT $\xi = \frac{2e\tau_B^{ASOT}}{j_e \hbar} = 0.053 \pm 0.003$, where $e$ is the electron charge, $j_e$ is the electric current density and $\hbar$ is the reduced Planck constant; this efficiency is comparable with the effective spin Hall angle of Pt ($0.056 \pm 0.005$) measured in a Pt/Py bilayer[7]. The corresponding ASOT conductivity for 32 nm Py is calculated as $\sigma^{ASOT} = \frac{2e}{\hbar}\frac{\tau_B^{ASOT}}{E} = \xi\sigma = 2300 \pm 115 \, \Omega^{-1}\text{cm}^{-1}$, where $E$ is the applied electric field. In Fig. 3d, the deviation of the ASOT-induced change in Kerr ellipticity from the model for the 4 nm Py sample can be accounted for if a 1% variation between $\tau_T^{ASOT}$ and $\tau_B^{ASOT}$ is assumed, which may be due to a slight difference in spin relaxation at the two interfaces (see Supplementary Information section S6 for further discussion).

---

* All the uncertainties in this letter are single standard deviation uncertainties. The principle source of uncertainty here is the fitting uncertainty, which is determined by a linear regression analysis by plotting the experimental data as a function of the simulation results.

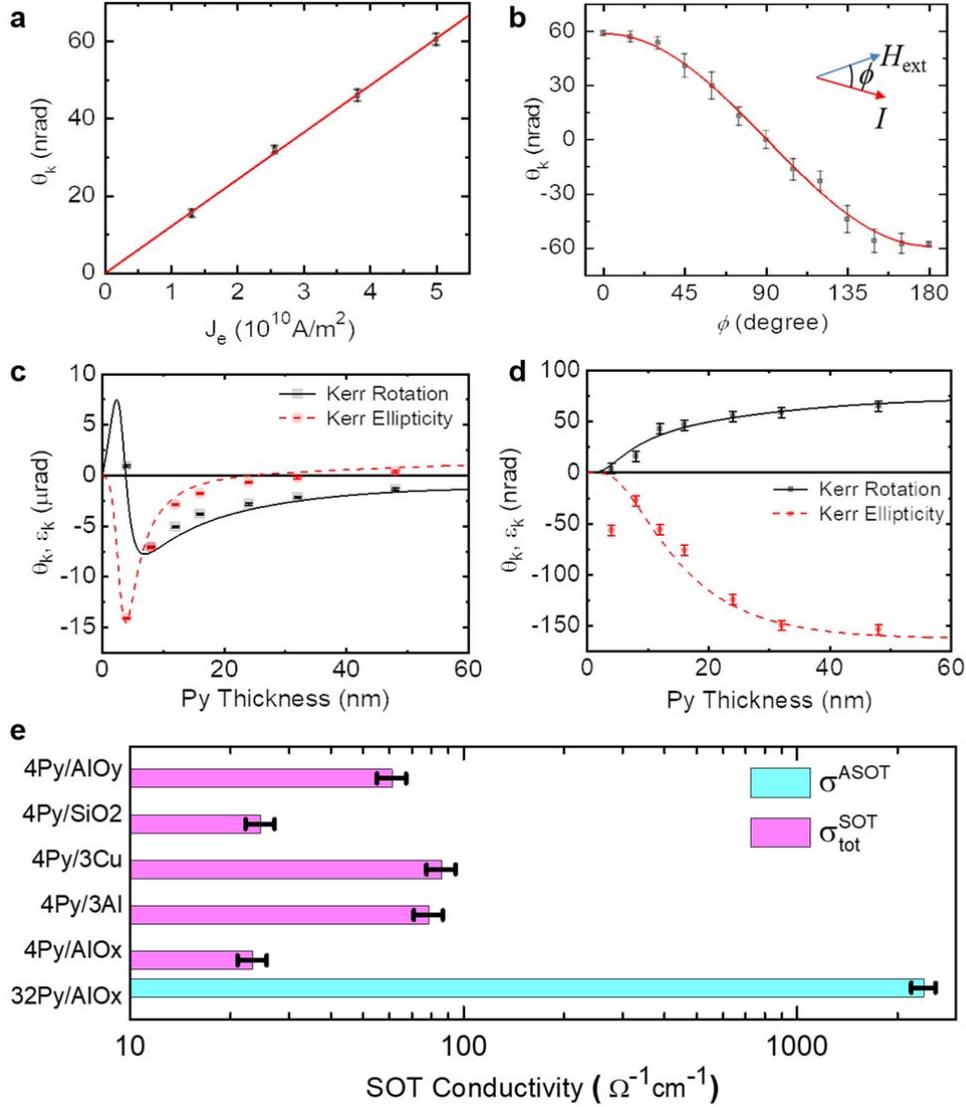

Figure 3 **Dependence of ASOT on current density, angle, thickness and the interface.** Kerr rotation change as a function of **a**, current density and **b**, the angle between current direction and magnetization. Kerr rotation (experimental, black squares; fit, black solid line) and ellipticity change (experimental, red circles; fit, red dashed line) **c**, due to the calibration field, and **d**, due to ASOT. **e**, Comparison between total SOT conductivities ($\sigma_{tot}^{SOT}$) measured for 4 nm Py with different capping layers, and the bottom-surface ASOT conductivity ($\sigma^{ASOT}$) of 32 nm Py. Error bars indicate single standard deviation uncertainties. In all these samples, the other side of the Py is in contact with AlO$_x$.

Since ASOT results in magnetization changes near the surface, the extracted ASOT values may be influenced by spin-orbit interaction at the interface with the capping layer, such as Rashba-Edelstein spin-orbit coupling[21-23]. To determine the relative contribution of such interface effects, we compare the ASOT at the top surface of the AlO$_x$(3)/Py(32)/AlO$_x$(3) sample with the total spin-orbit torque (SOT) in a series of control samples, AlO$_x$(3)/Py(4)/Cap, where Cap is varied among AlO$_x$(3), AlO$_y$(3, different oxidation time), SiO$_2$(3), Cu(3)/SiO$_2$(3) and Al(3)/SiO$_2$(3). These capping layer materials are often assumed to have weak spin-orbit interaction due to their being light elements, but they will change the electrostatic properties of the top interface, thereby changing the interfacial Rashba spin-orbit coupling. The bottom surface is the same as for the 32 nm Py sample and thus any interfacial contribution from the bottom surface should have similar ASOT conductivity. Since Py is only 4 nm in these control samples (thinner than the exchange length), the magnetization uniformly responds to the total SOT, which is a sum of the ASOTs at the top and bottom surfaces $(\tau_T^{ASOT} + \tau_B^{ASOT})$. Should there be a significant interface-dependence of the ASOT, a large total SOT will be observed in some of these control samples with asymmetric interfaces. As shown in Fig. 3e, all samples exhibit total SOT conductivities $\sigma_{tot}^{SOT} = \frac{2e}{\hbar}(\tau_T^{ASOT} + \tau_B^{ASOT})/E$ of at most 4% of the bottom-surface ASOT conductivity of the 32 nm Py sample. This suggests that the top-surface ASOT, which varies less than 4% among Py with different capping layers, does not contain a substantial contribution from the interface of the Py with the capping layers.

The insensitivity of ASOT to the interface implies that it arises from the bulk spin-orbit interaction within the magnetic material. ASOT can be phenomenologically understood as the result of the TSHE – a flow of transversely polarized spin current generates ASOT by transferring spin angular momentum from one surface to the other. We evaluate the TSHE conductivity using linear response in the Kubo formalism in the clean limit using density functional theory[24] (see Supplementary Information section S7 for technical details). First-principles calculations for Ni, Fe and Co all show significant TSHE conductivities, summarized in Table 1. We also measure the ASOT conductivities of these materials experimentally, provided in Table 1. For comparison, we also calculate and measure the AHE conductivities for these materials. If the ASOT is only due to the TSHE from the intrinsic band structure, the calculated TSHE conductivity should match the measured ASOT conductivity. As shown in Table 1, the conductivities are similar in magnitude as those calculated, indicating that the intrinsic mechanism may significantly contribute to the ASOT. However, the signs for Fe and Co are opposite between measured and calculated values; this may be because that the intrinsic mechanism is not the sole source for ASOT, and that other mechanisms should be taken into account. By analogy with the AHE, we expect that extrinsic mechanisms such as skew scattering[10,25] can also contribute to generating transversely polarized spin current and hence ASOT (see Supplementary Information Fig. S8).

|  |  | Ni | Fe | Co |
|---|---|---|---|---|
| Calculation | Structure | FCC | BCC | HCP |
|  | AHE Conductivity | -1.3 | 0.72 | 0.45 |
|  | TSHE Conductivity | 3.92 | 1.05 | -0.24 |
| Experiment | Structure | FCC | BCC | HCP |
|  | Conductivity | 56 | 32 | 46 |
|  | AHE Conductivity | -0.5± 0.05 | 0.5± 0.05 | 0.3± 0.03 |
|  | ASOT Conductivity | 3.5 ± 0.1 | -1.0 ± 0.2 | 0.8 ± 0.5 |

Table 1. **Measured and calculated electrical, AHE and ASOT conductivities.** All values have units of $10^3\,\Omega^{-1}\text{cm}^{-1}$. All experimental data are extrapolated based on 40 nm sputtered polycrystalline films, sandwiched between two 3 nm $\text{AlO}_x$ layers. The positive sign for the ASOT conductivity corresponds to the scenario that if the applied electric field is in the x-direction, the generated spin current flowing in the z-direction has spin moment in the -y-direction. Under this choice, the spin Hall conductivity of Pt is positive.

Our results unambiguously reveal a large surface torque with properties that are largely independent of the interface consisting of light elements, indicating that a considerable amount of transversely polarized spin current can exist within a ferromagnetic metal. Interconversion between charge and transverse spins in magnetic multilayers have been actively studied very recently[26-29], and the spin-charge conversion has been attributed to an interfacial spin-orbit interaction[30]. Our results suggest that the bulk spin-orbit interaction within the ferromagnet should also be taken into account.

Although structural symmetry dictates that the total ASOT equals zero in an isolated magnetic layer with symmetric surfaces, such symmetry is likely broken in a magnetic multilayer with structurally and compositionally different interfaces, and where a neighboring layer can act as a strong spin sink, either by spin absorption in the bulk of an adjacent layer, or by spin memory loss at the interface. In either case, the net result is that some fraction of the transversely polarized spin current that is generated in the TSHE is asymmetrically dissipated into the lattice without affecting the magnetic layer. As a result, the torques acting on the top and bottom surfaces of the ferromagnet are no longer symmetric and a net ASOT is generated (see Supplementary Information Fig. S10). Therefore, we speculate that our work may also challenge the fundamental understanding of SOT in ferromagnet/nonmagnet multilayers, in which the possibility of an inherent source of SOT acting on the FM had generally been overlooked.

# Methods

## Sample Fabrication

The samples used in this study are fabricated via magnetron sputtering. The $AlO_x$ layers are made by depositing 2 nm Al film and subsequent oxidization in an oxygen plasma.

## MOKE Measurement of ASOT

The MOKE measurements are performed with a lock-in balanced detection system[20], which is illustrated in Supplementary Information Fig. S3. An alternating current with frequency 20.15 kHz is applied through the patterned sample and the ASOT-induced MOKE response at the same frequency is measured. We use a Ti:sapphire mode-locked laser with ≈100 fs pulses at 80 MHz repetition rate with center wavelength 780 nm; the detectors used are slow relative to the repetition rate, so the measured signals are averaged over the pulses. The laser beam is focused by a 10x microscope objective into a spot of ~4 $\mu$m diameter. Laser power below 4 mW is used to avoid significant heating effect. To eliminate the quadratic MOKE contribution, the average is taken of the signals for incident laser polarizations of 45° and 135° with respect to the magnetization[20]. A combination of a second half-wave plate and a Wollaston prism is used to analyze the Kerr rotation signal. For Kerr ellipticity measurements, a quarter-wave plate is inserted before the half-wave plate.

## Fitting of the thickness-dependent MOKE signal

In the simulations, the magnetic film is discretized into many sublayers of thickness 0.4 nm. By assuming equal and opposite ASOT $\tau_T^{ASOT} = -\tau_B^{ASOT}$ at the top and bottom of each sublayer, we calculate the resultant out-of-plane magnetization using numerical methods (see Supplementary Information section S4). For calibration, a constant out-of-plane calibration field $h_{Cal}$ is applied to all sublayers, and the out-of-plane magnetization is calculated using the same numerical methods. Based on the calculated out-of-plane magnetization distribution, the polar MOKE response is determined using the propagation matrix method and taking into account multiple reflections (see Supplementary Information section S5). The above processes provide linear relationships between $\tau_T^{ASOT}$ and $h_{Cal}$ with the predicted MOKE response for various film thicknesses. In fitting the thickness-dependent MOKE response for calibration, shown in Fig. 3a, all fitting parameters are measured by other techniques. The good agreement corroborates our numerical model. In the fitting of the thickness-dependent MOKE response due to ASOT, shown in Fig. 3b, we assume $\tau_T^{ASOT}$ is the same for all film thicknesses under the same current density. All other fitting parameters are the same as those used in fitting to the calibration result. The good agreement shown in Fig. 3b confirms our assumption that $\tau_\text{T}^\text{ASOT}$ depends on the current density.

**Acknowledgements**

The work done at the University of Denver is partially supported by the PROF and by the National Science Foundation under Grant Number ECCS-1738679. W.W., D.G.C. and V.O.L. acknowledge support from the NSF-MRSEC under Award Number DMR-1720633. T.W., Y.W, and J.Q.X acknowledge support from NSF under Award Number DMR-1505192. V.P.A. acknowledges support under the Cooperative Research Agreement between the University of Maryland and the National Institute of Standards and Technology Center for Nanoscale Science and Technology, Award 70NANB14H209, through the University of Maryland. We would also like to thank Mark Stiles and Emilie Jue for critical reading of the manuscript, and Xiao Li for illuminating discussions.

**Contributions**

X.F. and H.O. conceived the idea; X.F., W.W., and T.W. designed the experiments; T.W. fabricated the sample, T.W., Y.W., W.W., A.R., A.D., T.J.S., D.B. and B.L.Z. patterned and characterized the samples; W.W. performed the MOKE measurements; W.W., X.F., V.O.L., D.G.C. and J.Q.X. analyzed the data; V.P.A. and P.M.H. carried out the first-principles calculations. X.F., W.W., V.O.L., V.P.A. and P.M.H. prepared the manuscript; all authors commented on the manuscript.

**Competing interests**

The authors declare no competing financial interest.